\documentclass[english,aps,prb,amsmath,twocolumn,preprintnumber,superscriptaddress,showpacs,notitlepage,nofootinbib]{revtex4-1}

\makeatletter
\usepackage{mathptmx,newtxtext,newtxmath,xspace}
\usepackage{amsbsy,bm,bbold}
\usepackage{graphicx,color,xcolor,epsfig,rotate}
\usepackage{gensymb}
\usepackage[colorlinks=true,
            linkcolor=blue,
            urlcolor=blue,
            citecolor=blue]{hyperref}
\usepackage{soul}

\allowdisplaybreaks[2]

\makeatother

\usepackage{babel}
\begin{document}
\title{Theory of the Kitaev model in a [111] magnetic field}
\author{Shang-Shun~Zhang}
\affiliation{Department of Physics and Astronomy, The University of Tennessee,
Knoxville, Tennessee 37996, USA}
\affiliation{School of Physics and Astronomy, University of Minnesota, Minneapolis, Minnesota 55455, USA}
\author{G{\'a}bor~B.~Hal{\'a}sz}
\affiliation{Materials Science and Technology Division, Oak Ridge
National Laboratory, Oak Ridge, Tennessee 37831, USA}
\affiliation{Quantum Science Center, Oak Ridge, Tennessee 37831, USA}
\author{Cristian~D.~Batista}
\affiliation{Department of Physics and Astronomy, The University of Tennessee,
Knoxville, Tennessee 37996, USA}
\affiliation{Neutron Scattering Division and Shull-Wollan Center, Oak Ridge
National Laboratory, Oak Ridge, Tennessee 37831, USA}

\begin{abstract}
{\bf Recent numerical studies indicate that the antiferromagnetic Kitaev honeycomb lattice model undergoes a magnetic-field-induced quantum phase transition into a new spin-liquid phase. This intermediate-field phase has been previously characterized as a gapless spin liquid. By implementing a recently developed variational approach based on the exact fractionalized excitations of the zero-field model, we demonstrate that the field-induced spin liquid is gapped and belongs to Kitaev's 16-fold way. Specifically, the low-field non-Abelian liquid with Chern number $C=\pm 1$ transitions into an Abelian liquid with $C=\pm 4$. The critical field and the field-dependent behaviors of key physical quantities are in good quantitative agreement with  published numerical results. Furthermore, we derive an effective field theory for the field-induced critical point which readily explains the ostensibly gapless nature of the intermediate-field spin liquid.}
\end{abstract}

\maketitle

\noindent{
\textbf{Introduction.} The exactly solvable Kitaev model on the honeycomb lattice~\cite{kitaev2006anyons} has deepened our insight into quantum spin liquids and helped us in identifying strongly spin-orbit-coupled $4d$ and $5d$ materials that may host these exotic quantum phases of matter~\cite{jackeli2009mott, chaloupka2010kitaev}. Indeed, recent years have seen a flurry of such ``Kitaev materials'' in which the microscopic spin Hamiltonian is believed to approximately realize the Kitaev honeycomb model~\cite{rau2016spin, trebst2017kitaev, hermanns2018physics, takagi2019concept}. The most famous ones include the honeycomb iridates, Na$_2$IrO$_3$~\cite{singh2010antiferromagnetic, liu2011longrange, choi2012spin, ye2012direct, comin2012novel, chun2015direct}, $\alpha$-Li$_2$IrO$_3$~\cite{singh2012relevance, williams2016incommensurate}, and H$_3$LiIr$_2$O$_6$~\cite{kitagawa2018spin}, as well as the honeycomb halide $\alpha$-RuCl$_3$~\cite{plumb2014spin, sandilands2015scattering, sears2015magnetic, majumder2015anisotropic, johnson2015monoclinic, sandilands2016spin, banerjee2016proximate, banerjee2017neutron, do2017majorana}. }

While most of these materials are magnetically ordered at the lowest temperatures, the zigzag magnetic order in $\alpha$-RuCl$_3$ can be suppressed with an in-plane magnetic field~\cite{Kubota_2015, Majumder_2015, Johnson_2015, leahy2017anomalous, Sears_2017, wolter2017field, baek2017evidence, Banerjee_2018, hentrich2018unusual, Nejc_2018}. Also, there are some experimental indications for an intermediate-field spin-liquid phase between the low-field magnetically ordered phase and the high-field spin-polarized phase. Most importantly, a recent experimental work~\cite{Kasahara_2018b} reported a half-integer-quantized thermal Hall conductivity in the intermediate-field regime just beyond the transition out of zigzag order. Though the exact nature of this regime is still an open question, the ongoing experimental efforts reveal the importance of precisely characterizing field-induced spin-liquid phases.

Motivated in large part by the intriguing experimental observations, the behavior of the Kitaev model in a magnetic field has been extensively studied~\cite{Janssen_2019} by various approaches, including exact diagonalization~\cite{Hickey_2019, kaib2019kitaev, zhu2018robust}, density-matrix renormalization group (DMRG)~\cite{zhu2018robust, Patel_2019}, infinite DMRG (iDMRG)~\cite{Gohlke_2017}, tensor-network methods~\cite{lee2020magnetic}, continuous-time quantum Monte Carlo techniques~\cite{yoshitake2020majorana}, and slave-particle mean-field theories~\cite{berke2020field}. These approaches all give consistent results. While the ferromagnetic Kitaev model has a single transition into a polarized phase, the antiferromagnetic Kitaev model includes a new intermediate-field spin liquid between the low-field non-Abelian spin liquid~\cite{kitaev2006anyons} and the high-field polarized phase.

In this work, we implement a novel variational approach~\cite{Zhang_2021} to investigate the ground-state phase diagram of the antiferromagnetic Kitaev model in a magnetic field parallel to the [111] direction. This approach is based on the exact fractionalized Majorana-fermion (``spinon'') and gauge-flux (``vison'') excitations of the pure Kitaev model at zero field~\cite{kitaev2006anyons}. It accounts for two effects of the magnetic field: the renormalization of the Majorana dispersion through a hybridization with pairs of fluxes (see Fig.~\ref{fig1}~\textbf{a}) and the finite dispersion acquired by the flux pairs themselves (see Fig.~\ref{fig1}~\textbf{b}). Remarkably, we find a continuous quantum phase transition, induced by a softening of a hybridized excitation, at a critical field $h_c \simeq 0.50$, which is very close to the critical field $h_c \simeq 0.44$ reported by a recent iDMRG study~\cite{Gohlke_2017}. The critical point signals the transition of the non-Abelian spin liquid~\cite{kitaev2006anyons} with Chern number $C=\pm 1$  into an Abelian spin liquid with $C=\pm 4$. The predicted field dependence of the flux expectation value and the second derivative of the ground-state energy is also in good quantitative agreement with the iDMRG results. Moreover, the effective field theory of the quantum critical point, as derived from the microscopic Hamiltonian, predicts a low-energy ring of gapped excitations in momentum space, which is difficult to be distinguished from a gapless Fermi surface in finite systems. We conjecture that this is the main reason why previous works~\cite{Hickey_2019, zhu2018robust, Patel_2019} characterized the phase at $h \gtrsim h_c$ as a gapless spin liquid.

\vspace{0.3cm}
\noindent{
\textbf{Model.} 
We consider the \textit{antiferromagnetic} Kitaev model~\cite{kitaev2006anyons} in an external magnetic field along the [$111$] direction,
\begin{eqnarray}
\mathcal{H} = \sum_{\alpha}\sum_{\mathbf{r}\in A} \sigma^{\alpha}_{\mathbf{r}} \sigma^{\alpha}_{\mathbf{r}+\hat{\mathbf{r}}_{\alpha}}
 + h \sum_{\mathbf{r}} (\sigma^{x }_{\mathbf{r}}+\sigma^y_{\mathbf{r}}+\sigma^z_{\mathbf{r}}), \label{eq-model}
\end{eqnarray}
where $h$ is the magnetic field (in units of the Kitaev energy) and $\hat{\mathbf{r}}_{\alpha}$ is the nearest-neighbor vector from an $A$ site to a $B$ site along an $\alpha$ bond (see Fig.~\ref{fig1}). For the exactly solvable Kitaev model in the $h=0$ limit, the low-energy spectrum comprises gapless matter fermions (i.e., spinons) with a single Dirac cone and gapped dispersionless $\mathbb{Z}_2$ gauge fluxes. These elementary excitations are described in terms of four Majorana fermions $c_{\mathbf{r}}^{\phantom{x}}$ and ${b}_{\mathbf{r}}^{\alpha}$ with $\alpha=x,y,z$ at each site $\mathbf{r}$, where $c_{\mathbf{r}}^{\phantom{x}}$ are the matter fermions, and ${b}_{\mathbf{r}}^{\alpha}$ are bond fermions associated with the $\mathbb{Z}_2$ gauge field $u_{\mathbf{r},\mathbf{r}+\hat{\mathbf{r}}_{\alpha}}^{\alpha} \equiv i b_{\mathbf{r}}^{\alpha} b_{\mathbf{r}+\hat{\mathbf{r}}_{\alpha}}^{\alpha} = \pm 1$. The gauge fields are conserved bond variables that commute with each other; their product around any plaquette $p$ (see Fig.~\ref{fig1}~$\mathbf{a}$) is gauge invariant and expressible in terms of the physical spins:
\begin{equation}
W_p = u_{12}^z u_{32}^x u_{34}^y u_{54}^z u_{56}^x u_{16}^y =
\sigma_1^x \sigma_2^y \sigma_3^z \sigma_4^x \sigma_5^y \sigma_6^z.
\label{eq-mod-W}
\end{equation}
Thus, $W_p = \pm 1$ can be identified as static $\mathbb{Z}_2$ gauge fluxes. In each flux sector, $\{ W_p = \pm 1 \}$, represented with an appropriate gauge-field configuration, $\{u_{\mathbf{r},\mathbf{r}+\hat{\mathbf{r}}_{\alpha}}^{\alpha} = \pm 1\}$, the zero-field model then reduces to a quadratic fermion problem.

\begin{figure}[t!]
\centering
\includegraphics[width=0.9\columnwidth]{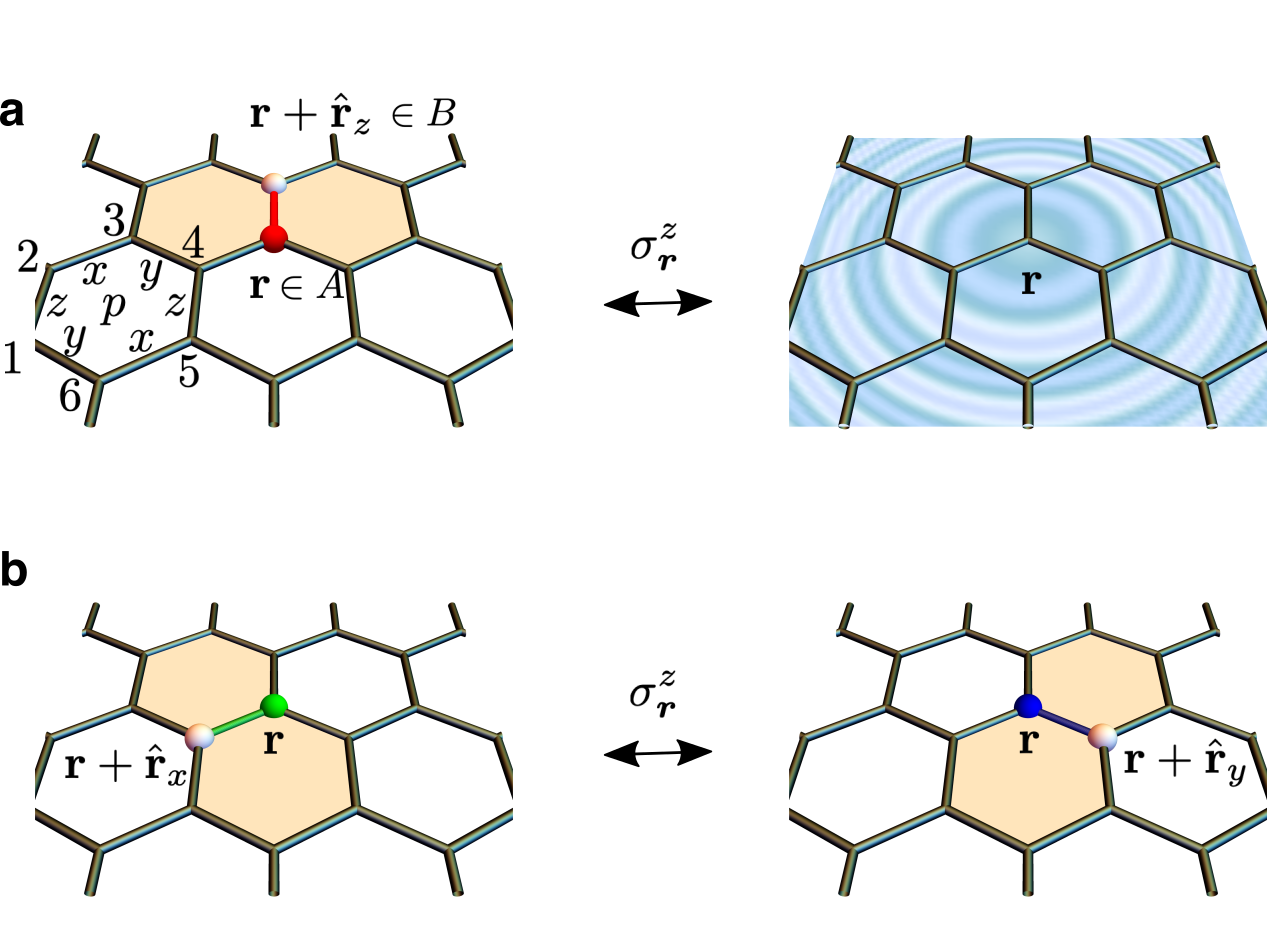}
\caption{\textbf{Effect of the magnetic field.} \textbf{a} Hybridyzation between a flux pair and a fermion. \textbf{b} Hopping of a flux pair between two neighboring bonds. Definitions of the plaquettes $p$, the two sublattices $A$ and $B$, and the nearest-neighbor bond vectors $\hat{\mathbf{r}}_{x,y,z}$ are also shown. \label{fig1}}
\end{figure}

While the model in Eq.~\eqref{eq-model} is not exactly solvable for a finite field, we can derive a low-energy effective model by projecting $\mathcal{H}$ into the low-energy sector of the pure Kitaev model (corresponding to $h=0$) generated by single matter-fermion and/or flux-pair excitations~\cite{Zhang_2021}. We focus on flux pairs because, unlike single fluxes, they are coherent fermionic quasiparticles~\cite{Zhang_2021} and can readily hybridize with matter fermions (see Fig.~\ref{fig1}~$\mathbf{a}$). The resulting low-energy Hamiltonian reads
\begin{eqnarray}
\tilde{\mathcal{H}} & = & \sum_{\alpha}\sum_{\mathbf{r}\in A}i c_{\mathbf{r}}^{\phantom{x}} c_{\mathbf{r}+\hat{\mathbf{r}}_{\alpha}}^{\phantom{x}}+\Delta_{\chi} \sum_{\alpha}\sum_{\mathbf{r}\in A}(\tilde{\chi}_{\mathbf{r}}^{\alpha})^{\dag}(\tilde{\chi}_{\mathbf{r}}^{\alpha}) \nonumber \\
& - & i h q\sum_{\alpha,\beta} \epsilon_{\alpha\beta} \left\{ \sum_{\mathbf{r}\in A}(\tilde{\chi}_{\mathbf{r}}^{\alpha})^{\dag}\tilde{\chi}_{\mathbf{r}}^{\beta}+\sum_{\mathbf{r}\in B} (\tilde{\chi}_{\mathbf{r}-\hat{\mathbf{r}}_{\alpha}}^{\alpha})^{\dag}\tilde{\chi}_{\mathbf{r}-\hat{\mathbf{r}}_{\beta}}^{\beta}\right\} \nonumber \\
& + &  h\sum_{\alpha}\sum_{\mathbf{R}}p_{\mathbf{R},\alpha}^{\phantom{x}}\left[\sum_{\mathbf{r}\in A}i\tilde{b}_{\mathbf{r}}^{\alpha}c_{\mathbf{r}+\mathbf{R}}^{\phantom{\alpha}}+\sum_{\mathbf{r}\in B}i\tilde{b}_{\mathbf{r}}^{\alpha}c_{\mathbf{r}-\mathbf{R}}^{\phantom{\alpha}}\right] ,\label{eq-H}
\end{eqnarray}
where $A$ and $B$ are the two honeycomb sublattices, $\mathbf{R}$ is a general lattice vector, and $\epsilon_{\alpha \beta} = \sum_{\gamma} \epsilon_{\alpha \beta \gamma}$ is an antisymmetric symbol based on the Levi-Civita symbol $\epsilon_{\alpha \beta \gamma}$. The first two terms correspond to the pure Kitaev model: the first term describes the matter fermions, and the second term accounts for the finite energy ($\Delta_{\chi} \simeq 0.26$) of a flux pair. The dressed bond-fermion operator, $(\tilde{\chi}_{\mathbf{r}\in A}^{\alpha})^{\dagger}= \frac{1}{2} (\tilde{b}_{\mathbf{r}}^{\alpha}-i\tilde{b}_{\mathbf{r}+\hat{\mathbf{r}}_{\alpha}}^{\alpha})$, simultaneously creates a flux pair on the $\alpha$ bond connected to the site $\mathbf{r} \in A$ and distorts the matter-fermion state: $(\tilde{\chi}_{\mathbf{r}}^{\alpha})^{\dagger}\rvert \omega\rangle \otimes \rvert 0 \rangle = \, \rvert \phi_{\bm r}^{\alpha} \rangle \otimes \rvert \chi_{\mathbf{r}}^{\alpha} \rangle$, where $\rvert \omega \rangle$ and $\rvert \phi_{\bm r}^{\alpha} \rangle$ are the matter-fermion vacua of the gauge-field configurations $\rvert 0 \rangle$ and $\rvert \chi_{\mathbf{r}}^{\alpha} \rangle$ which correspond to the flux-free sector and the single-flux-pair sector, respectively. Formally, $\rvert \chi_{\mathbf{r}}^{\alpha} \rangle = (\chi_{\mathbf{r}}^{\alpha})^{\dagger} \rvert 0 \rangle$, where $(\chi_{\mathbf{r}}^{\alpha})^{\dagger}= \frac{1}{2} (b_{\mathbf{r}}^{\alpha}-i b_{\mathbf{r}+\hat{\mathbf{r}}_{\alpha}}^{\alpha})$ is a bare bond-fermion operator, while $\rvert 0 \rangle$ is the vacuum of these bare bond fermions that has $u_{\mathbf{r},\mathbf{r}+\hat{\mathbf{r}}_{\alpha}}^{\alpha} = -1$ for all bonds. We note that the dressed bond fermions $\tilde{\chi}_{\mathbf{r}}^{\alpha}$ have the same projective symmetries as the bare bond fermions ${\chi}_{\mathbf{r}}^{\alpha}$. The remaining two terms in Eq.~\eqref{eq-H} account for the finite magnetic field: the third term describes flux-pair hopping (see Fig.~\ref{fig1}~\textbf{b}), and the fourth term corresponds to the most general symmetry-allowed hybridization between a flux pair and a matter fermion (see Fig.~\ref{fig1}~\textbf{a}). We note that $q$ and $p_{\mathbf{R},\alpha}$ are dimensionless coefficients and that some $p_{\mathbf{R},\alpha}$ are identical due to threefold rotation symmetry acting simultaneously in real space and spin space. The effective Hamiltonian $\tilde{\mathcal{H}}$ is quadratic and can be straightforwardly diagonalized in momentum space:
\begin{eqnarray}
\tilde{\mathcal{H}} & = & \sum_{\mathbf{k}} \left[ i \lambda_{\mathbf{k}}^{\phantom{\dagger}} C_{\mathbf{k},A}^{\dagger} C_{\mathbf{k},B}^{\phantom{\dagger}} + \mathrm{H.c.} \right] \nonumber \\
& & + \sum_{\mathbf{k},\alpha,\beta} \left\{ \Delta_{\chi} \delta_{\alpha \beta} - i h q \epsilon_{\alpha \beta} \left[ 1+e^{i \mathbf{k}\cdot (\hat{\mathbf{r}}_{\alpha}-\hat{\mathbf{r}}_{\beta})} \right] \right\} (\tilde{X}_{\mathbf{k}}^{\alpha})^{\dag}(\tilde{X}_{\mathbf{k}}^{\beta}) \nonumber \\ 
&  & + {h\over \sqrt{2}}\sum_{\mathbf{k},\alpha} \Big\{ i P_{\mathbf{k},\alpha}^{\phantom{x}} \left[ \tilde{X}_{-\mathbf{k}}^{\alpha} + (\tilde{X}_{\mathbf{k}}^{\alpha})^{\dagger} \right] C_{\mathbf{k},A}^{\phantom{x}} \nonumber \\ 
&  &  + P_{-\mathbf{k},\alpha}^{\phantom{x}} \, e^{i \mathbf{k}\cdot \hat{\mathbf{r}}_{\alpha} } \left[ \tilde{X}_{-\mathbf{k}}^{\alpha} - (\tilde{X}_{\mathbf{k}}^{\alpha})^{\dagger} \right] C_{\mathbf{k},B}^{\phantom{x}} + \mathrm{H.c.} \Big\}, \label{eq-H-k}
\end{eqnarray}
where $\lambda_{\mathbf{k}}=\sum_{\alpha} e^{i\mathbf{k}\cdot \hat{\mathbf{r}}_{\alpha}}$ and $P_{\mathbf{k},\alpha} = \sum_{\mathbf{R}}p_{\mathbf{R},\alpha} \, e^{i \mathbf{k}\cdot \mathbf{R}}$, while
\begin{eqnarray}
C_{\mathbf{k},\nu} = \frac{1}{\sqrt{2N}} \sum_{\mathbf{r} \in \nu} c_{\mathbf{r}} e^{-i \mathbf{k}\cdot \mathbf{r}}, \;\;\;\;\;\;
\tilde{X}_{\mathbf{k}}^{\alpha} = \frac{1}{\sqrt{N}}  \sum_{\mathbf{r} \in A} \tilde{\chi}_{\mathbf{r}}^{\alpha} e^{-i \mathbf{k}\cdot \mathbf{r}} \,\,
\end{eqnarray}
are momentum-space matter and bond fermions in terms of the sublattice index $\nu = A,B$ and the system size $N$. By considering the matrix elements of the Zeeman term $\propto h$ in Eq.~\eqref{eq-model} within the low-energy sector of the pure Kitaev model~\cite{Zhang_2021}, we relate the dimensionless parameters in Eq.~\eqref{eq-H-k} to matter-fermion matrix elements of this exactly solvable model~\cite{supp}:
\begin{eqnarray}
q &=& \langle\phi_{\mathbf{0}}^{\beta} \rvert (1+ic_{\mathbf{0}}^{\phantom{x}} c_{\hat{\mathbf{r}}_{\alpha}}^{\phantom{x}}) \rvert \phi_{\mathbf{0}}^{\gamma} \rangle, \quad \alpha \neq \beta \neq \gamma, \nonumber \\
P_{\mathbf{k},\alpha}^{\phantom{x}} &=& \langle \phi_{\mathbf{0}}^{\alpha}  \rvert \omega\rangle + {1\over 2} \sum_{\mathbf{k}^{\prime}} (1-e^{-i\mathbf{k}^{\prime}\cdot\hat{\mathbf{r}}_{\alpha} + i \varphi_{\mathbf{k}^{\prime}} }) \langle \phi_{\mathbf{0}}^{\alpha}  \rvert \psi_{\mathbf{k}^{\prime}}^{\dagger} \psi_{\mathbf{k}}^{\dagger}  \rvert \omega\rangle, \quad
\label{eq-p}
\end{eqnarray}
where $\mathbf{r} = \mathbf{0}$ is an $A$ site, while $\psi_{\mathbf{k}}=(C_{\mathbf{k},A} + i e^{i \varphi_{\mathbf{k}}} C_{\mathbf{k},B})/\sqrt{2}$ in terms of $e^{i \varphi_{\mathbf{k}}}=\lambda_{\mathbf{k}}/\rvert \lambda_{\mathbf{k}} \rvert$ are the matter fermions diagonalizing the flux-free sector of the pure Kitaev model. For a finite honeycomb lattice with $N = 121\times121$ unit cells, we numerically find $q\simeq 0.0494$ and $P_{\mathbf{0},\alpha} \simeq 0.722$.


}

\begin{figure}[t!]
\centering
\includegraphics[width=1\columnwidth]{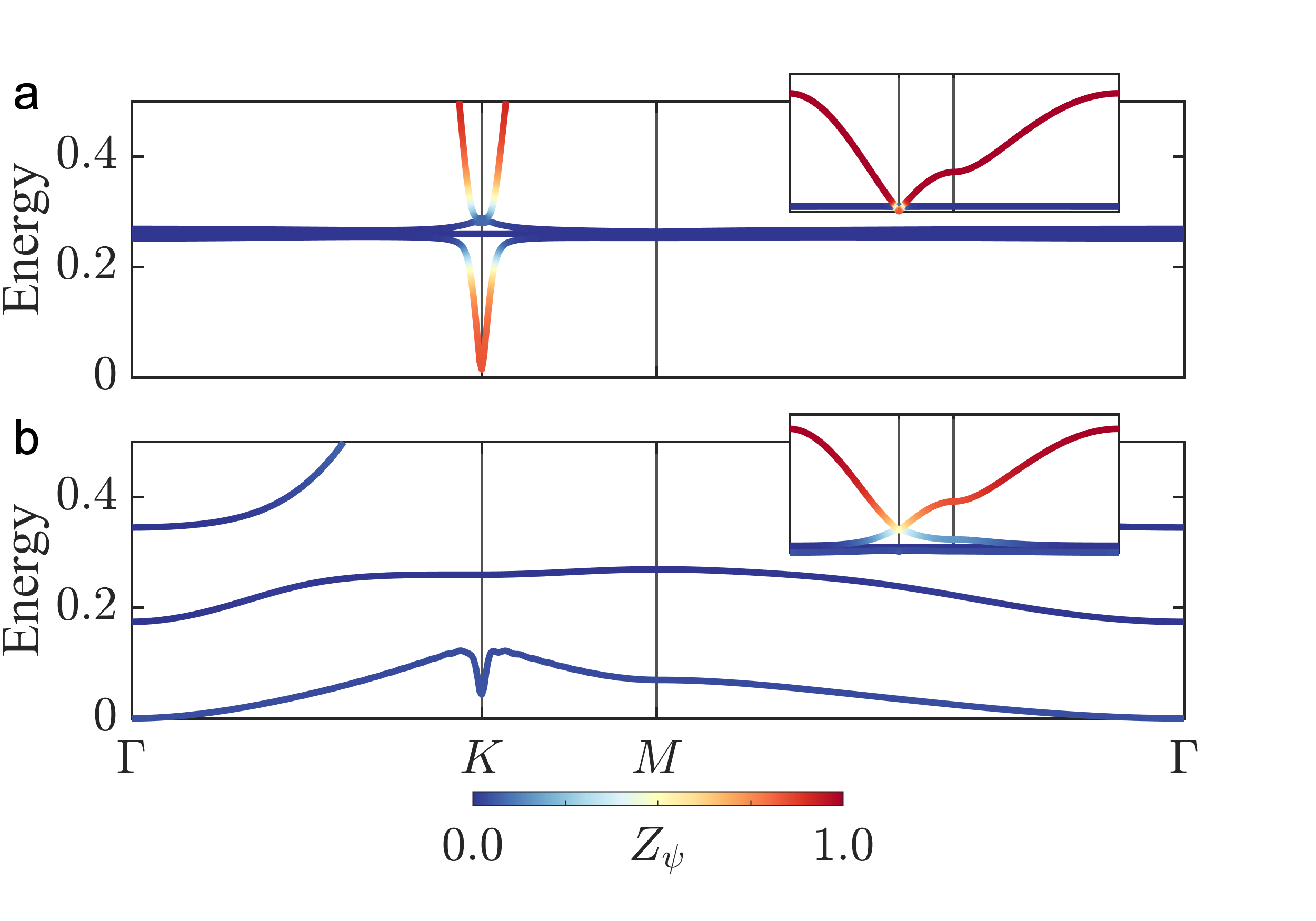}
\caption{\textbf{Low-energy spectrum of the effective Hamiltonian.} The fermion dispersions correspond to $h=0.05$ in \textbf{a} and $h=h_c\simeq 0.50$ in \textbf{b}. The color scale shows the  matter-fermion weight, $0 < Z_{\psi} < 1$, of the given fermion eigenmode; red (blue) color indicates predominantly matter-fermion (bond-fermion) character. The insets show the spectrum over the full energy range. Note that a hybridization decay length, $\xi = 25$, is used to regularize the K-point behavior.\cite{footnote}  \label{fig2}}
\end{figure}

\vspace{0.3cm}
\noindent{
\textbf{Results.}
We study the low-energy effective model in Eq.~\eqref{eq-H} as a function of the magnetic field $h$. At zero field, the spectrum coincides with that of the pure Kitaev model and contains one dispersive matter-fermion band as well as the three flat bond-fermion bands (see Fig.~\ref{fig2}~\textbf{a}). For a small field, $h \ll \Delta_{\chi}$, the hybridization between these four bands gives rise to a finite energy gap, $\Delta_K (h) \propto h^3$, at the K point of the Brillouin zone (BZ). The slow field dependence of $\Delta_K (h)$, which is expected from a perturbative argument by Kitaev~\cite{kitaev2006anyons}, explains why the global minimum of the band structure remains at the K point up to a large field, $h_0 \simeq 0.46$. As shown in Fig.~\ref{fig3}~\textbf{a}, the global minimum switches from the K point to the $\Gamma$ point at $h = h_0$, and the corresponding gap, $\Delta_{\Gamma} (h)$, closes at a slightly larger field, $h_c \simeq 0.50$ (see Fig.~\ref{fig2}~\textbf{b}). Since the little group of the $\Gamma$ point includes the threefold rotation C$_3$, the fermion eigenmodes at the $\Gamma$ point can be classified according to their C$_3$ eigenvalues. The natural bond-fermion modes, corresponding to C$_3$ eigenvalues $1$ and $e^{\mp 2\pi i/3}$, respectively, are then
\begin{align}
\tilde{X}^{0}_{\bm 0} &= \left(\tilde{X}^x_{\bm 0} + \tilde{X}^y_{\bm 0} + \tilde{X}^z_{\bm 0}\right)/\sqrt{3}, \nonumber \\ 
\tilde{X}^{\pm}_{\bm 0} &= \left(\tilde{X}^x_{\bm 0} + e^{\pm 2\pi i/3} \tilde{X}^y_{\bm 0}+ e^{\mp 2\pi i/3} \tilde{X}^z_{\bm 0}\right)/\sqrt{3}.
\end{align}
Since the matter-fermion mode $\psi_{\bm 0}$ is invariant under C$_{3}$, it can only hybridize with the bond-fermion mode $\tilde{X}^{0}_{\bm 0}$. At the critical field, $h_c = 3\sqrt{\Delta_{\chi}/2} \, (\sum_{\alpha}P_{{\bm 0},\alpha})^{-1} \simeq 0.50$, one of the resulting hybridized eigenmodes is gapless. In contrast, there is a higher critical field, $h_c' = \Delta_{\chi}/(2\sqrt{3}q) \simeq 1.52$ (not shown in Fig.~\ref{fig3}), at which the pure bond-fermion eigenmode $\tilde{X}^{+}_{\bm 0}$ has vanishing energy. We note that a complete diagonalization over the full BZ reveals yet another critical point at $h_c'' \simeq 1.0$ due to the softening of a hybridized mode at the M point. We emphasize, however, that the effective model is no longer expected to be valid when $h$ is significantly larger than $h_c$.

\begin{figure}[t!]
\centering
\includegraphics[width=0.85\columnwidth]{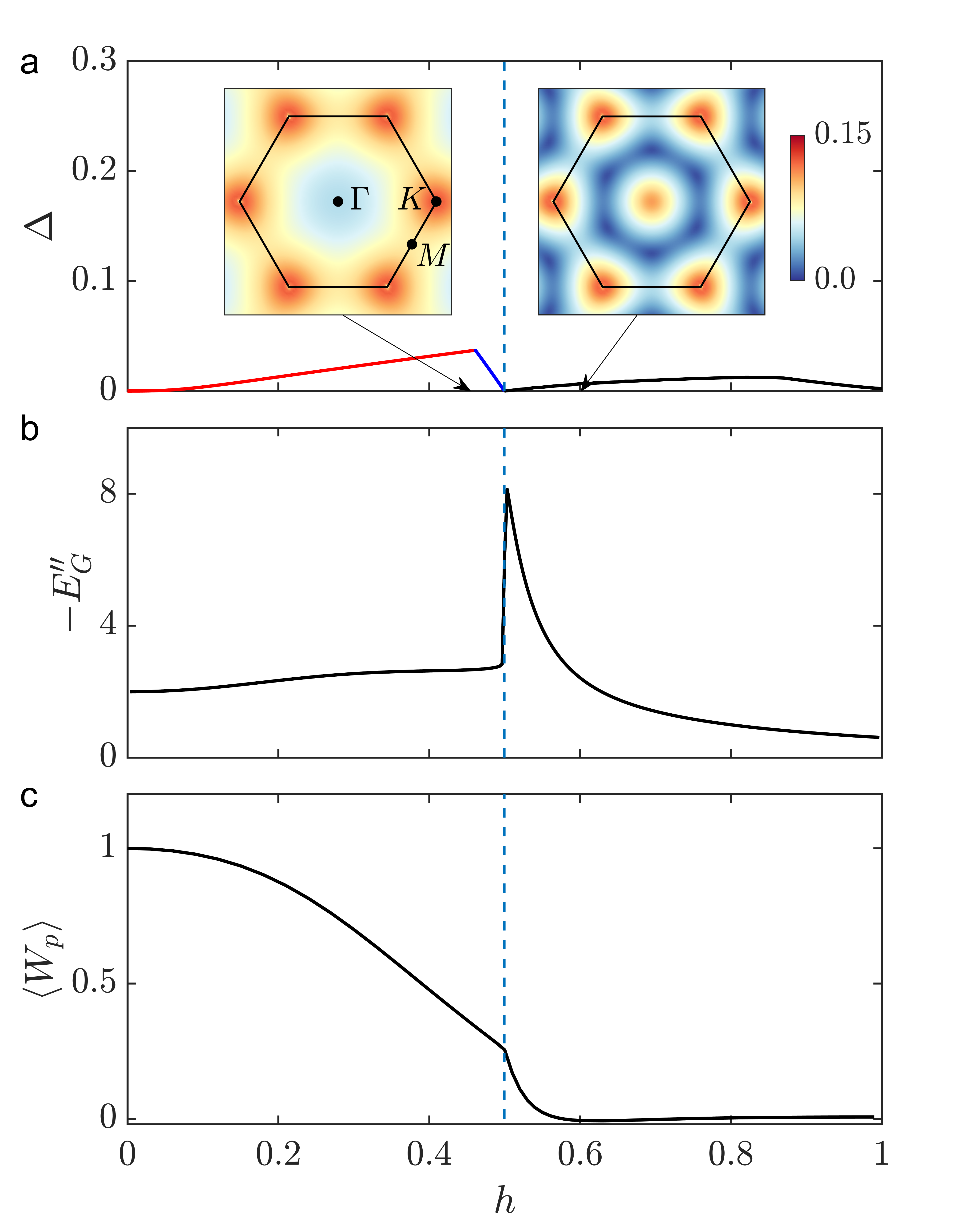}
\caption{\textbf{Field dependence of key physical quantities.} \textbf{a} Overall energy gap. The insets show the dispersion of the low-energy fermion eigenmode on both sides of the phase transition. The red (blue) line corresponds to the gap $\Delta_{K}$ ($\Delta_{\Gamma}$), while the black line corresponds to the gap at the six wave vectors ${\bm Q}_j$, the corners of the blue hexagon in the right-hand-side inset. \textbf{b} Second derivative of the ground-state energy. \textbf{c} Expectation value of the $\mathbb{Z}_2$ gauge flux. \label{fig3}}
\end{figure}

Figure~\ref{fig3}~\textbf{a} shows the overall energy gap as a function of the magnetic field $h$. As expected, the gap is proportional to $h^3$ at the smallest fields, $h \ll \Delta_{\chi}$. Just below $h_c$, the global minimum of the excitation spectrum switches from the K point to the $\Gamma$ point, and the gap vanishes at $h_c \simeq 0.50$~\cite{Gohlke_2017,zhu2018robust,Hickey_2019}. Importantly, the zero-energy mode at $h = h_c$ has dominant bond-fermion character with a large bond-fermion weight $6/(6+\Delta_{\chi}) \simeq 0.96$ (see also Fig.~\ref{fig2}~\textbf{b}), which is consistent with the numerical closing of the vison gap in the specific heat~\cite{Hickey_2019}. In contrast, the gap reopens for $h \gtrsim h_c$, which appears to be in contradiction with the same numerical results and the corresponding conjecture of a gapless U(1) spin liquid at intermediate fields. However, our analytic approach can also explain the numerical similarity between the gapped spin liquid at $h \gtrsim h_c$ and a gapless spin liquid with a circular spinon Fermi surface. Indeed, as we explain below, the phase transition at $h=h_c$ gives rise to a low-energy ring at $h \gtrsim h_c$ (see the inset of Fig.~\ref{fig3}~\textbf{a}) which expands from the $\Gamma$ point and corresponds to a small energy gap $\propto (h - h_c)^{3/2}$. This low-energy ring naturally explains the large low-energy density of states found by exact diagonalization~\cite{zhu2018robust,Hickey_2019}. The emergence of the low-energy ring and the nature of the $h \gtrsim h_c$ phase are explained in the next section, where we derive an effective field theory to describe the continuous topological phase transition at $h=h_c$.

Figures~\ref{fig3}~\textbf{b} and \textbf{c} plot the second derivative of the ground-state energy,  $E_G^{\prime \prime} = d^2E_G/dh^2$, and the expectation value of the $\mathbb{Z}_2$ gauge flux, $\langle W_p \rangle$, against the magnetic field.
As we explain below, the discontinuity of $E_G^{\prime \prime}$ at $h=h_c$ is a generic property of the corresponding phase transition. This discontinuity leads to a peak in $E_G^{\prime \prime}$ at $h=h_c$, which is qualitatively and quantitatively consistent with the iDMRG results~\cite{Gohlke_2017}. We note that our result for $\langle W_p \rangle$~\cite{supp} (see Fig.~\ref{fig3}~\textbf{c}) is also consistent with iDMRG.

Since the effective Hamiltonian in Eq.~\eqref{eq-H} is not expected to be valid when $ \langle W_p \rangle \ll 1$, we focus on the lowest-field phase transition at $h=h_c$ throughout the rest of this work. Remarkably, the critical field $h_c \simeq 0.50$ is only 10\% higher than the corresponding iDMRG result, $h_c \simeq 0.44$~\cite{Gohlke_2017}. Furthermore, the slight overestimation of $h_c$ is not surprising because the inclusion of higher-energy ($E \simeq 2 \Delta_{\chi}$) states with four fluxes and one matter fermion would lead to a reduction of $h_c$. Finally, at $h=h_c$, the dynamical spin structure factor from iDMRG indicates that the spin excitation gap closes at the $\Gamma$ point, which is in agreement with our results for the critical field.\footnote{We note that a spin excitation fractionalizes into a pair of fermion excitations. Since the fermions are gapless at the $\Gamma$ point, a pair of gapless fermions has a vanishing total momentum, corresponding to a vanishing spin gap at the $\Gamma$ point.} These similarities between the iDMRG results and those obtained from our effective Hamiltonian $\tilde{\mathcal{H}}$ indicate that our variational low-energy manifold captures the essence of the phase transition at $h=h_c$ and the new spin-liquid phase at $h \gtrsim h_c$.
}

\vspace{0.3cm}
\noindent{
\textbf{Field theory of topological phase transition.}
In the vicinity of the critical field, $h \simeq h_c \simeq 0.50$, the low-energy fermion eigenmodes belong to the trivial representation of C$_3$, and the long-wavelength limit of $\tilde{\mathcal{H}}$, corresponding to the region around the $\Gamma$ point, can be written as
\begin{equation}
\tilde{\mathcal{H}}_{\rm eff}=\sum_{{\bm{k}}} f_{{\bm{k}}}^{\dagger}[\beta_{{\bm{k}}}^{x}\tau_{x}+\beta_{{\bm{k}}}^{y}\tau_{y}+\beta_{{\bm{k}}}^{z}\tau_{z}]f_{{\bm{k}}}^{\phantom{\dagger}},
\label{Heff}
\end{equation}
where $\tau_{x,y,z}$ are the Pauli matrices, and $f_{{\bm{k}}}=(f_{1,{\bm{k}}},f_{2,{\bm{k}}})^T$ is a two-component fermionic operator corresponding to the two zero-energy modes of $\tilde{{\cal H}}$ at the critical field:
\begin{eqnarray}
f^{\phantom{\dagger}}_{1,{\bm k}} = \sqrt{6 \over 6+\Delta_{\chi}} \tilde{X}_{{\bm k}}^0 - i \sqrt{\Delta_{\chi}\over 6+\Delta_{\chi}} \psi_{{\bm k}}^{\phantom{0}} , \ \ \ \ f^{\phantom{\dagger}}_{2,{\bm k}} = f^{\dagger}_{1,-{\bm k}}.
\end{eqnarray}
The coefficients $\beta_{{\bm{k}}}^{x,y,z}$ in Eq.~\eqref{Heff} must be $C_3$ invariant real polynomials. Up to cubic order in $\bm{k}$, there are only four such polynomials: the trivial polynomial $1$, the quadratic polynomial $k^2 = k_x^2+k_y^2$, and the cubic polynomials $g_{{\bm{k}}}^{x}=k_{x}(3k_{y}^{2}-k_{x}^{2})$ and $g_{{\bm{k}}}^{y}=k_{y}(3k_{x}^{2}-k_{y}^{2})$. Moreover, the particle-hole symmetry of the original Hamiltonian $\mathcal{H}$ dictates that $\tilde{\mathcal{H}}_{\rm eff}$ must remain invariant under $f_{{\bm{k}}}^{\phantom{\dagger}}\rightarrow\tau_{x}(f_{-{\bm{k}}}^{\dagger})^{T}$, implying that the polynomials $\beta_{{\bm{k}}}^{\mu}$ must satisfy the following relationships:
\begin{equation}
\beta_{{\bm{k}}}^{x}=-\beta_{-{\bm{k}}}^{x},\ \ \beta_{{\bm{k}}}^{y}=-\beta_{-{\bm{k}}}^{y},\ \ \beta_{{\bm{k}}}^{z}=\beta_{-{\bm{k}}}^{z}.\label{eq:even_odd}
\end{equation}
These symmetry considerations then lead to the general forms
\begin{eqnarray}
\beta_{{\bm{k}}}^{z} &=& c_{0}+c_{z} k^2, \nonumber \\
\beta_{\bm{k}}^{\eta}&=&\sum_{\nu=x,y}c_{\eta \nu}g_{\bm{k}}^{\nu}, \quad \eta=x,y,
\end{eqnarray}
where $c_0$, $c_z$, and $c_{\eta \nu}$ are, in general, functions of $h$. Since the phase transition at $h=h_c$ is driven by a sign change in $c_0$, we assume that $c_z$ and $c_{\eta \nu}$ are constants, while we write $c_0 = c_0' (h-h_c)$ with a constant $c_0'$. Starting from Eqs.~\eqref{eq-H-k} and \eqref{eq-p}, and defining all lengths in units of the lattice vector (i.e., the distance between two neighboring $A$ sites), the constants are derived to be\cite{supp} $c_0' \simeq -1.00$, $c_z \simeq 0.0125$, $c_{xx} \simeq -0.00268$, $c_{yy} \simeq -0.00088$, and $c_{xy} = c_{yx} = 0$. Then, using Eq.~\eqref{Heff}, the fermion dispersion is given by
\begin{equation}
\omega_{{\bm{k}}}=\sqrt{\left(\beta_{{\bm{k}}}^{x}\right)^{2}+\left(\beta_{{\bm{k}}}^{y}\right)^{2}+\left(\beta_{{\bm{k}}}^{z}\right)^{2}}
\end{equation} 
and becomes gapless at $\bm{k} = \bm{0}$ for $h=h_c$.
For $h<h_c$, the dispersion is dominated by the function $\beta_{{\bm{k}}}^{z}$ and is largely quadratic: $\omega_{{\bm{k}}} \simeq \rvert c_{0}' \rvert (h_c-h)+c_{z}k^{2}$. In contrast, for $h>h_c$, the function $\beta_{{\bm{k}}}^{z}$ vanishes for $\rvert{\bm{k}}\rvert=\sqrt{\rvert c_{0}' \rvert (h-h_c)/c_{z}}$. Thus, along this ring of radius $\rvert{\bm{k}}\rvert$, the energy gap is determined by the small cubic contributions from $\beta_{{\bm{k}}}^{x,y}$ and has a slow field dependence: $\Delta \propto (h-h_c)^{3/2}$. The net result is a ring of low-energy fermions around the $\Gamma$ point (see the inset of Fig.~\ref{fig3} \textbf{a}). 

The effective field theory in Eq.~\eqref{Heff} describes a continuous topological phase transition. The phases on both sides of the transition belong to Kitaev's 16-fold way~\cite{kitaev2006anyons} and are characterized by the fermion Chern number. The contribution from the low-energy fermions to this Chern number is given by~\cite{Zhang_2020}
\begin{eqnarray}
C &=& {1\over 4\pi} \int d{\bm k} \, {\bm d}_{\bm k}\cdot [\partial_{k_x} {\bm d}_{\bm k} \times \partial_{k_y} {\bm d}_{\bm k}], 
\end{eqnarray}
where ${\bm d}_{\bm k}={\bm \beta}_{\bm k}/\rvert {\bm \beta}_{\bm k} \rvert$ and ${\bm \beta}_{\bm k}=(\beta_{\bm k}^x,\beta_{\bm k}^y,\beta_{\bm k}^z)$. 
Geometrically, $C$ is simply the skyrmion  number of the vector field ${\bm d}_{\bm k}$. Figure~\ref{fig4} depicts the vector field ${\bm d}_{\bm k}$ around the $\Gamma$ point on both sides of the phase transition at $h=h_c$. While the field configuration is topologically trivial for $h<h_c$, it includes six merons (three skyrmions) for $h>h_c$. The corresponding change in the Chern number, $\Delta C=3$, is then a generic property of the phase transition described by  $\tilde{\mathcal{H}}_{\rm eff}$. To understand the emergence of the six merons around the $\Gamma$ point, we first note that ${\beta}_{\bm k}^{\eta} \propto \text{Im}(k_+^3 e^{-i\phi_{\eta}})$ with $k_{+} = k_x +i k_y$ and $\phi_{\eta}=\arctan(c_{\eta x}/c_{\eta y})$. Each function ${\beta}_{\bm k}^{\eta}$ (with $\eta = x,y$) possesses three nodal lines corresponding to $k_y/k_x = \tan(\phi_{\eta}/3 + \varphi)$ with $\varphi = 0, \pi/3, 2\pi/3$. Ignoring the $\beta_{\bm k}^y$ function, the low-energy spectrum then contains six Dirac nodes ${\bm Q}_j$ (with $j=1,2,...,6$) at the intersections of the nodal lines of $\beta_{\bm k}^x$ and the ring of radius $\rvert{\bm{k}}\rvert=\sqrt{\rvert c_{0}' \rvert (h-h_c)/c_{z}}$. The vorticity of the vector field ${\bm d}_{\bm k}$ around each Dirac node ${\bm Q}_j$ is $(-1)^j$. Assuming $\phi_x\neq \phi_y$ (which is true in our case), the finite value of $\beta_{{\bm Q}_j}^y \propto (-1)^j$ generates a mass term for each Dirac node in such a way that the Dirac nodes all give identical contributions ($+1/2$ each or $-1/2$ each) to the change in the Chern number. The net change in the Chern number is then
\begin{equation}
\Delta C = 3 \, \mathrm{sgn} \left[\det\hat{{\cal C}}\right], \quad \hat{{\cal C}}=\left(\begin{array}{ccc}
c_{xx} & c_{xy} & 0\\
c_{yx} & c_{yy} & 0 \\
0 & 0 & c_z \end{array}\right).
\end{equation}
Using the constants $c_z$ and $c_{\eta \nu}$ given above, we obtain $\Delta C=3$ at the critical field $h=h_c$. Since the low-field phase at $h<h_c$ is well known~\cite{kitaev2006anyons} to have Chern number $1$, we conclude that the higher-field phase at $h \gtrsim h_c$ has Chern number $4$.

\begin{figure}[t!]
\centering
\includegraphics[width=1\columnwidth]{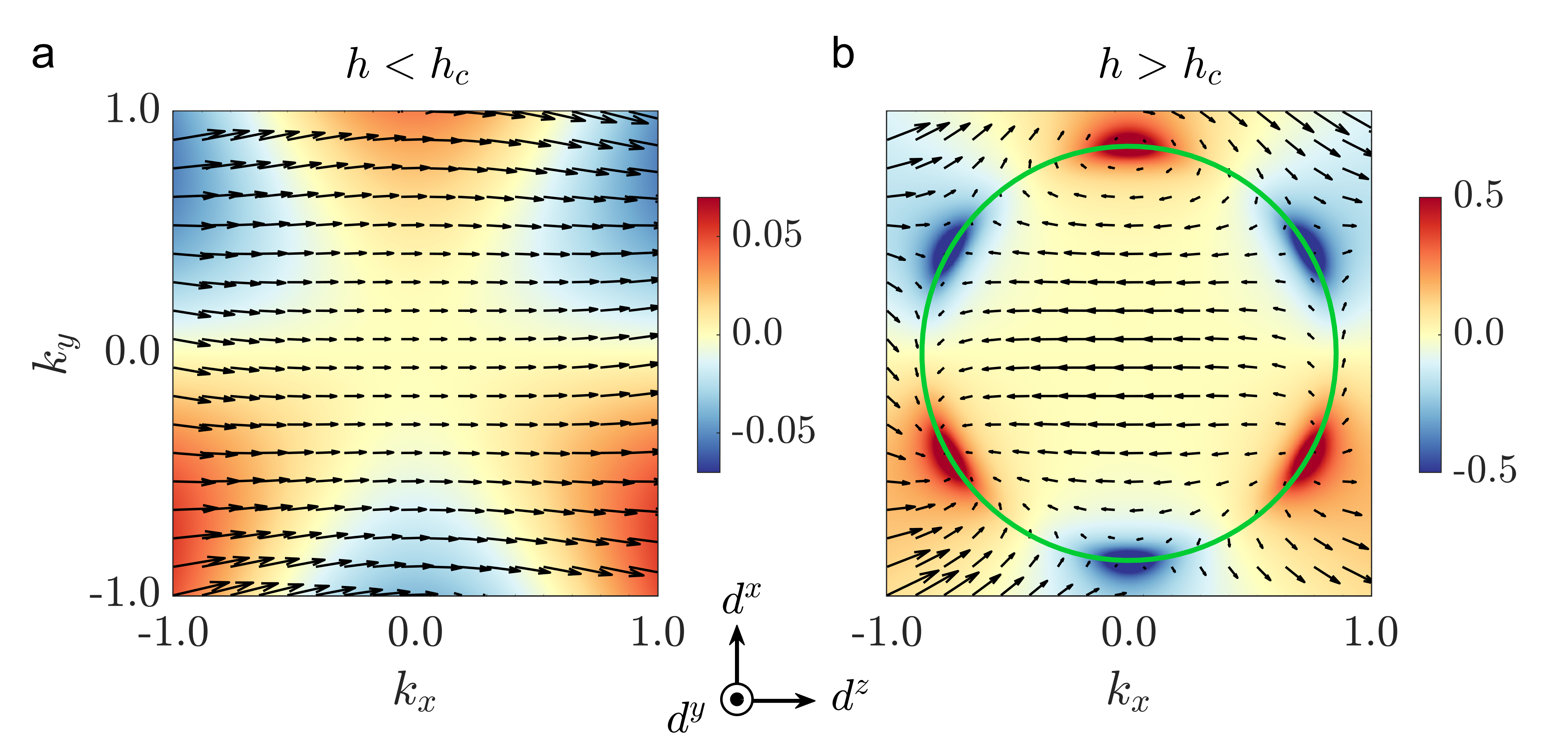}
\caption{\textbf{Topological phase transition.} Configuration of the unit-vector field ${\bm d}_{\bm k}$ on the two sides of the phase transition. The color scale shows the component $d_{\bm k}^y$, while the black arrows represent the components $(d_{\bm k}^{z},d_{\bm k}^{x})$. The green circle marks the low-energy ring.
\label{fig4} }
\end{figure}

We next consider the second derivative of the ground-state energy $E_G^{\prime\prime}$ with respect to the magnetic field $h$. The universal critical behavior at $h=h_c$ is determined by the low-energy modes $\rvert {\bm k}\rvert \leq \Lambda$, where the cutoff $\Lambda$ can be made arbitrarily small (corresponding to an infrared singularity). 
While the contribution of these modes to $E_G^{\prime\prime}$ is $\propto \Lambda^2$ for ${h\rightarrow h_c^-}$, it is an ${\cal O}(1)$ constant for ${h\rightarrow h_c^+}$. In particular, there is a contribution from the neighborhood of the low-energy ring at $h \gtrsim h_c$ which is independent of the cutoff $\Lambda$. Therefore, we obtain a discontinuity in $E_G^{\prime\prime}$ at the critical field~\cite{supp}:
\begin{equation}
\Delta E_{G}^{\prime \prime} = \lim _{h \rightarrow h_c^{-}} E_{G}^{\prime \prime}-\lim _{h \rightarrow h_c^{+}} E_{G}^{\prime \prime}=\frac{\sqrt{3}(c_{0}')^{2}}{8 \pi c_{z}}.    
\end{equation}
Remarkably, this discontinuity in $E_G^{\prime\prime}$, as shown in Fig.~\ref{fig3}~\textbf{b}, is entirely determined by two coefficients of the effective field theory. From the constants $c_0'$ and $c_z$ given above, it is found to be $\Delta E_{G}^{\prime \prime} \simeq 5.5$, which is consistent with the corresponding result for a finite lattice (see Fig.~\ref{fig3}~\textbf{b}). The quantitative agreement between this value and the one obtained from iDMRG~\cite{Gohlke_2017} indicates that the effective field theory at $h=h_c$ is both qualitatively correct and quantitatively accurate. 
}


\vspace{0.3cm}
\noindent{
\textbf{Discussion.}
Our simple and accurate variational approach to extended Kitaev models~\cite{Zhang_2021} indicates that the antiferromagnetic (AFM) Kitaev model undergoes a continuous quantum phase transition driven by a magnetic field parallel to the [111] direction. According to this approach, the new phase, which has been reported in previous numerical works~\cite{Hickey_2019, kaib2019kitaev, zhu2018robust, Patel_2019, Gohlke_2017}, is a gapped chiral spin liquid with a ring of low-energy excitations. Due to its large low-energy density of states, it is difficult for numerical simulations to distinguish this low-energy ring from a gapless Fermi surface. In contrast to the non-Abelian low-field phase, the new phase at higher fields possesses Abelian topological order with four anyon types: $1$ (vacuum), $\varepsilon$ (fermion), as well as $e$ and $m$ (vortices). The two phases can then be distinguished numerically by computing the entanglement spectrum~\cite{yao2010entanglement} or the topological entanglement entropy for a bipartition of an infinite cylinder~\cite{zhang2012quasiparticle, cincio2013characterizing, zaletel2013topological}, readily available in iDMRG~\cite{Gohlke_2017}. The addition of irrelevant Hamiltonian terms that increase the gap in the higher-field phase without generating new phase transitions may help in numerically testing our predictions. }


We also emphasize that our approach straightforwardly generalizes to the ferromagnetic (FM) Kitaev model. In this case, the first term in Eq.~\eqref{eq-H} has a negative sign, and the flux-pair-hopping parameter in Eq.~\eqref{eq-p} is found to be $q\simeq 1.35$, i.e., about $30$ times larger than for the AFM Kitaev model. Therefore, the lowest-field phase transition is driven by a softening of a pure flux-pair mode and happens at a much smaller critical field, $h_c' = \Delta_{\chi} / (2\sqrt{3}q) \simeq 0.056$. The strong asymmetry between the FM and AFM Kitaev models is due to opposite (constructive and destructive) interference effects between the two processes contributing to flux-pair hopping~\cite{Zhang_2021}. We note that this asymmetry is not apparent in the simplified perturbative analysis of Ref.~\onlinecite{kitaev2006anyons} because it neglects the energy dispersions of the intermediate states. Finally, we remark that our results for the FM Kitaev model are also consistent with numerical studies that report a single first-order transition into a trivial polarized phase at a critical field $h_p \simeq 0.028$~\cite{Gohlke_2017}. At this first-order phase transition, corresponding to $h_p \lesssim h_c'$, the fluxes suddenly proliferate and confine all fractionalized excitations.

\vspace{0.3cm}

\noindent{Dated: \today}

\bibliographystyle{naturemag}
\bibliography{paper}

\vspace{0.3cm}




\vspace{0.3cm}

\noindent{\textbf{Data availability}}

\noindent{The data that support the findings of this study are available from the corresponding authors upon reasonable request.}

\vspace{0.3cm}

\noindent{\textbf{Code availability}}

\noindent{The codes that support the findings of this study are available from the corresponding authors upon reasonable request.}

\vspace{0.3cm}

\noindent{\textbf{Acknowledgements}}

\noindent{We thank Matthias Gohlke for useful discussions. 
S-S.~Z. and C.~D.~B. are supported by funding from the Lincoln Chair of Excellence in Physics. 
G.~B.~H.~was supported by the U.S. Department of Energy, Office of Science, National Quantum Information Science Research Centers.}

\vspace{0.3cm}

\noindent{\textbf{Author contributions}}

\noindent{All authors made significant contributions to the manuscript.}

\vspace{0.3cm}

\noindent{\textbf{Competing interests}}

\noindent{The authors declare no competing interests.}

\clearpage

\widetext

\setcounter{figure}{0}
\renewcommand{\thefigure}{S\arabic{figure}}
\setcounter{equation}{0}
\renewcommand{\theequation}{S\arabic{equation}}


\begin{center}
\textbf{\large Supplementary Information for ``Theory of the Kitaev model in a [111] magnetic field''}
\end{center}

\section{Dimensionless parameters of the effective Hamiltonian}

Here we derive the momentum-space hybridization parameters, $P_{\mathbf{k},\alpha} = \sum_{\mathbf{R}} p_{\mathbf{R}, \alpha} \, e^{i \mathbf{k} \cdot \mathbf{R}}$, and the flux-pair hopping parameter $q$ in Eq.~(6) of the main text. We first consider the hybridization parameters. Using the momentum-space complex matter fermions diagonalizing the flux-free sector of the pure Kitaev model, 
\begin{equation}
\psi_{\mathbf{k}}=\frac{1}{\sqrt{2}}\left(C_{\mathbf{k},A} + i e^{i \varphi_{\mathbf{k}}} C_{\mathbf{k},B}\right)=\frac{1}{2\sqrt{N}}\sum_{\mathbf{r}\in A}c_{\mathbf{r}}e^{-i\mathbf{k}\cdot\mathbf{r}}+\frac{i}{2\sqrt{N}}\sum_{\mathbf{r}\in B}c_{\mathbf{r}}e^{-i\mathbf{k}\cdot\mathbf{r}+i\varphi_{\mathbf{k}}},\label{eq-psi}
\end{equation}
where $e^{i\varphi_{\mathbf{k}}}=\lambda_{\mathbf{k}}/|\lambda_{\mathbf{k}}|$ and $\lambda_{\mathbf{k}}=\sum_{\alpha}e^{i\mathbf{k}\cdot\hat{\mathbf{r}}_{\alpha}}$, the hybridization term in Eq.~(4) of the main text can then be written as
\begin{eqnarray}
\tilde{\mathcal{H}}_{p} & = & \frac{h}{\sqrt{N}}\sum_{\alpha}\sum_{\mathbf{k}}\sum_{\mathbf{r}\in A}\left[i(\tilde{\chi}_{\mathbf{r}}^{\alpha})^{\dag}\psi_{\mathbf{k}}^{\phantom{\dag}}e^{i\mathbf{k}\cdot\mathbf{r}}  \left( P_{\mathbf{k},\alpha}^{\phantom{x}}  + P_{-\mathbf{k},\alpha}^{\phantom{x}}  e^{ i\mathbf{k}\cdot\hat{\mathbf{r}}_{\alpha}-i\varphi_{\mathbf{k}}}\right)+\mathrm{H.c.}\right]\nonumber \\
&  & +\frac{h}{\sqrt{N}}\sum_{\alpha}\sum_{\mathbf{k}}\sum_{\mathbf{r}\in A}\left[i(\tilde{\chi}_{\mathbf{r}}^{\alpha})^{\dag}\psi_{-\mathbf{k}}^{\dag}e^{i\mathbf{k}\cdot\mathbf{r}} \left( P_{\mathbf{k},\alpha}^{\phantom{x}}  - P_{-\mathbf{k},\alpha}^{\phantom{x}} e^{ i \mathbf{k}\cdot\hat{\mathbf{r}}_{\alpha}-i\varphi_{\mathbf{k}}}\right)+\mathrm{H.c.}\right].
 \label{eq-H-p}
\end{eqnarray}
The two terms in Eq.~(\ref{eq-H-p}) can be matched with the microscopic model by considering the matrix elements of the bare Zeeman term, $\mathcal{H}_h = h \sum_{\alpha} \sum_{\mathbf{r}} i b_{\mathbf{r}}^{\alpha} c_{\mathbf{r}}^{\phantom{x}}$, between appropriate states of the pure Kitaev model: the ground state, $| \Omega \rangle = | \omega \rangle \otimes | 0 \rangle$, the matter-fermion eigenstate, $| \psi_{\mathbf{k}}^{\phantom{\dag}} \rangle = (\psi_{\mathbf{k}}^{\dag} | \omega \rangle) \otimes | 0 \rangle$, the flux-pair eigenstate, $| \tilde{\chi}_{\mathbf{r}}^{\alpha} \rangle = | \phi_{\mathbf{r}}^{\alpha} \rangle \otimes [(\chi_{\mathbf{r}}^{\alpha})^{\dag} | 0 \rangle]$, and the approximate eigenstate containing both a matter fermion and a flux pair, $| \tilde{\chi}_{\mathbf{r}}^{\alpha} \psi_{-\mathbf{k}}^{\phantom{\dag}} \rangle = (\psi_{-\mathbf{k}}^{\dag} | \phi_{\mathbf{r}}^{\alpha} \rangle) \otimes [(\chi_{\mathbf{r}}^{\alpha})^{\dag} | 0 \rangle]$, where $\chi_{\mathbf{r} \in A}^{\alpha} = \frac{1}{2} (b_{\mathbf{r}}^{\alpha} + i b_{\mathbf{r} + \hat{\mathbf{r}}_{\alpha}}^{\alpha})$ are the bond fermions of the pure Kitaev model, $| 0 \rangle$ is the vacuum of these bond fermions, $| \omega \rangle$ is the matter-fermion ground state of the flux-free sector, and $| \phi_{\mathbf{r}}^{\alpha} \rangle$ is the matter-fermion ground state of the flux sector with a single flux pair around the $\alpha$ bond connecting the sites $\mathbf{r} \in A$ and $(\mathbf{r} + \hat{\mathbf{r}}_{\alpha}) \in B$. According to Eq.~(\ref{eq-H-p}), these matrix elements must be equal to
\begin{eqnarray}
\langle \tilde{\chi}_{\mathbf{r}}^{\alpha} | \mathcal{H}_h | \psi_{\mathbf{k}}^{\phantom{\dag}} \rangle &=& \frac{i h} {\sqrt{N}} \, e^{i \mathbf{k} \cdot \mathbf{r}} \left( P_{\mathbf{k}, \alpha}^{\phantom{x}} + P_{-\mathbf{k}, \alpha}^{\phantom{x}} e^{i \mathbf{k} \cdot \hat{\mathbf{r}}_{\alpha} - i \varphi_{\mathbf{k}}} \right), \nonumber \\
\langle \tilde{\chi}_{\mathbf{r}}^{\alpha} \psi_{-\mathbf{k}}^{\phantom{\dag}} | \mathcal{H}_h | \Omega \rangle &=& -\frac{i h} {\sqrt{N}} \, e^{i \mathbf{k} \cdot \mathbf{r}} \left( P_{\mathbf{k}, \alpha}^{\phantom{x}} - P_{-\mathbf{k}, \alpha}^{\phantom{x}} e^{i \mathbf{k} \cdot \hat{\mathbf{r}}_{\alpha} - i \varphi_{\mathbf{k}}} \right), \label{eq-H-h}
\end{eqnarray}
and, therefore, the hybridization parameters $P_{\mathbf{k}, \alpha}^{\phantom{x}}$ are found to be
\begin{eqnarray}
P_{\mathbf{k}, \alpha}^{\phantom{x}} &=& -\frac{i \sqrt{N}} {2h} \, e^{-i \mathbf{k} \cdot \mathbf{r}} \left[ \langle \tilde{\chi}_{\mathbf{r}}^{\alpha} | \mathcal{H}_h | \psi_{\mathbf{k}}^{\phantom{\dag}} \rangle - \langle \tilde{\chi}_{\mathbf{r}}^{\alpha} \psi_{-\mathbf{k}}^{\phantom{\dag}} | \mathcal{H}_h | \Omega \rangle \right] \nonumber \\
&=& -\frac{i \sqrt{N}} {2} \, e^{-i \mathbf{k} \cdot \mathbf{r}} \left[ \langle \phi_{\mathbf{r}}^{\alpha} | \left( i c_{\mathbf{r}} - c_{\mathbf{r} + \hat{\mathbf{r}}_{\alpha}} \right) \psi_{\mathbf{k}}^{\dag} | \omega \rangle + \langle \phi_{\mathbf{r}}^{\alpha} | \psi_{-\mathbf{k}}^{\phantom{\dag}} \left( i c_{\mathbf{r}} - c_{\mathbf{r} + \hat{\mathbf{r}}_{\alpha}} \right) | \omega \rangle \right] \label{eq-p-1} \\
&=& \frac{\sqrt{N}} {2} \, e^{-i \mathbf{k} \cdot \mathbf{r}} \left[ \langle \phi_{\mathbf{r}}^{\alpha} | \left( c_{\mathbf{r}} + i c_{\mathbf{r} + \hat{\mathbf{r}}_{\alpha}} \right) \psi_{\mathbf{k}}^{\dag} | \omega \rangle + \langle \phi_{\mathbf{r}}^{\alpha} | \big\{ \psi_{-\mathbf{k}}^{\phantom{\dag}}, c_{\mathbf{r}} + i c_{\mathbf{r} + \hat{\mathbf{r}}_{\alpha}} \big\} | \omega \rangle - \langle \phi_{\mathbf{r}}^{\alpha} | \left( c_{\mathbf{r}} + i c_{\mathbf{r} + \hat{\mathbf{r}}_{\alpha}} \right) \psi_{-\mathbf{k}}^{\phantom{\dag}} | \omega \rangle \right], \nonumber
\end{eqnarray}
where $\{ a, b \} \equiv ab + ba$. Setting $\mathbf{r} = \mathbf{0}$ without loss of generality, these hybridization parameters then become
\begin{eqnarray}
P_{\mathbf{k}, \alpha}^{\phantom{x}} &=& \frac{\sqrt{N}}{2} \langle \phi_{\mathbf{0}}^{\alpha} | \left( c_{\mathbf{0}} + i c_{\hat{\mathbf{r}}_{\alpha}} \right) \psi_{\mathbf{k}}^{\dag} | \omega \rangle + \frac{1}{2} \, \langle \phi_{\mathbf{0}}^{\alpha} | \omega \rangle \left( 1 - e^{i \mathbf{k} \cdot \hat{\mathbf{r}}_{\alpha} - i \varphi_{\mathbf{k}}} \right) \nonumber \\
&=& \frac{1}{2} \sum_{\mathbf{k}'} \langle \phi_{\mathbf{0}}^{\alpha} | \left[ \psi_{\mathbf{k}'}^{\phantom{\dag}} \big( 1 + e^{i \mathbf{k}' \cdot \hat{\mathbf{r}}_{\alpha} - i \varphi_{\mathbf{k}'}} \big) + \psi_{\mathbf{k}'}^{\dag} \big( 1 - e^{-i \mathbf{k}' \cdot \hat{\mathbf{r}}_{\alpha} + i \varphi_{\mathbf{k}'}} \big) \right] \psi_{\mathbf{k}}^{\dag} | \omega \rangle + \frac{1}{2} \, \langle \phi_{\mathbf{0}}^{\alpha} | \omega \rangle \left( 1 - e^{i \mathbf{k} \cdot \hat{\mathbf{r}}_{\alpha} - i \varphi_{\mathbf{k}}} \right) \label{eq-p-2} \\
&=& \langle \phi_{\mathbf{0}}^{\alpha} | \omega \rangle + \frac{1}{2} \sum_{\mathbf{k}'} \big( 1 - e^{-i \mathbf{k}' \cdot \hat{\mathbf{r}}_{\alpha} + i \varphi_{\mathbf{k}'}} \big) \langle \phi_{\mathbf{0}}^{\alpha} | \psi_{\mathbf{k}'}^{\dag} \psi_{\mathbf{k}}^{\dag} | \omega \rangle. \nonumber \end{eqnarray}
Importantly, each matter fermion $\psi_{\mathbf{k}}$ with a given energy $\varepsilon_{\mathbf{k}}$ belongs to a degenerate set of matter fermions whose momenta $\mathbf{k}$ are related by the various symmetries of the Kitaev model. From the perspective of these matter fermions, the presence of a flux pair is a local perturbation which only affects the two sites $\mathbf{0} \in A$ and $\hat{\mathbf{r}}_{\alpha} \in B$ connected by the corresponding $\alpha$ bond. Therefore, we can form appropriate linear combinations of the degenerate matter fermions such that only two linear combinations couple to the perturbation while the remaining ones have vanishing wave functions at both sites $\mathbf{0}$ and $\hat{\mathbf{r}}_{\alpha}$. Exploiting the residual inversion symmetry around the flux pair (i.e., the corresponding $\alpha$ bond), the natural choice for these two linear combinations is
\begin{equation}
\psi_{\varepsilon, \pm} = \mathcal{N}_{\varepsilon, \pm}^{-1} \sum_{\{\mathbf{k}\}_{\varepsilon}} \left( 1 \pm e^{i \mathbf{k} \cdot \hat{\mathbf{r}}_{\alpha} - i \varphi_{\mathbf{k}}} \right) \psi_{\mathbf{k}}, \qquad \mathcal{N}_{\varepsilon, \pm} = \sqrt{\sum_{\{\mathbf{k}\}_{\varepsilon}} \big| 1 \pm e^{i \mathbf{k} \cdot \hat{\mathbf{r}}_{\alpha} - i \varphi_{\mathbf{k}}} \big|^2}, \label{eq-psi-2}
\end{equation}
where $\{\mathbf{k}\}_{\varepsilon}$ is the set of all momenta $\mathbf{k}$ satisfying $\varepsilon_{\mathbf{k}} = \varepsilon$. Since the matter fermions $\psi_{\varepsilon, \pm}$ have eigenvalues $\pm i$ under the residual inversion symmetry (which acts projectively on the matter fermions), the two-fermion matrix element in Eq.~(\ref{eq-p-2}) then becomes
\begin{equation}
\langle \phi_{\mathbf{0}}^{\alpha} | \psi_{\mathbf{k}'}^{\dag} \psi_{\mathbf{k}}^{\dag} | \omega \rangle = \sum_{\pm} \mathcal{N}_{\varepsilon_{\mathbf{k}'}, \pm}^{-1} \, \mathcal{N}_{\varepsilon_{\mathbf{k}}, \mp}^{-1} \big( 1 \pm e^{i \mathbf{k}' \cdot \hat{\mathbf{r}}_{\alpha} - i \varphi_{\mathbf{k}'}} \big) \left( 1 \mp e^{i \mathbf{k} \cdot \hat{\mathbf{r}}_{\alpha} - i \varphi_{\mathbf{k}}} \right) \langle \phi_{\mathbf{0}}^{\alpha} | \psi_{\varepsilon_{\mathbf{k}'}, \pm}^{\dag} \, \psi_{\varepsilon_{\mathbf{k}}, \mp}^{\dag} | \omega \rangle. \label{eq-psi-3}
\end{equation}
Note that $\langle \phi_{\mathbf{0}}^{\alpha} | \psi_{\varepsilon_{\mathbf{k}'}, +}^{\dag} \, \psi_{\varepsilon_{\mathbf{k}}, +}^{\dag} | \omega \rangle = \langle \phi_{\mathbf{0}}^{\alpha} | \psi_{\varepsilon_{\mathbf{k}'}, -}^{\dag} \, \psi_{\varepsilon_{\mathbf{k}}, -}^{\dag} | \omega \rangle = 0$ because of inversion symmetry. Using the auxiliary identities
\begin{eqnarray}
\sum_{\{\mathbf{k}\}_{\varepsilon}} \left( 1 - e^{i \mathbf{k} \cdot \hat{\mathbf{r}}_{\alpha} - i \varphi_{\mathbf{k}}} \right) \left( 1 - e^{-i \mathbf{k} \cdot \hat{\mathbf{r}}_{\alpha} + i \varphi_{\mathbf{k}}} \right) &=& \mathcal{N}_{\varepsilon, -}^2, \nonumber \\
\sum_{\{\mathbf{k}\}_{\varepsilon}} \left( 1 + e^{i \mathbf{k} \cdot \hat{\mathbf{r}}_{\alpha} - i \varphi_{\mathbf{k}}} \right) \left( 1 - e^{-i \mathbf{k} \cdot \hat{\mathbf{r}}_{\alpha} + i \varphi_{\mathbf{k}}} \right) &=& 0, \label{eq-aux}
\end{eqnarray}
the momentum-space hybridization parameters in Eq.~(\ref{eq-p-2}) finally take the form
\begin{equation}
P_{\mathbf{k}, \alpha}^{\phantom{x}} = \langle \phi_{\mathbf{0}}^{\alpha} | \omega \rangle + \frac{1}{2} \left( 1 + e^{i \mathbf{k} \cdot \hat{\mathbf{r}}_{\alpha} - i \varphi_{\mathbf{k}}} \right) \mathcal{N}_{\varepsilon_{\mathbf{k}}, +}^{-1} \sum_{\varepsilon'} \mathcal{N}_{\varepsilon', -}^{\phantom{-1}} \langle \phi_{\mathbf{0}}^{\alpha} | \psi_{\varepsilon', -}^{\dag} \, \psi_{\varepsilon_{\mathbf{k}}, +}^{\dag} | \omega \rangle \equiv C + A (\varepsilon_{\mathbf{k}}) \left[ 1 + e^{i \mathbf{k} \cdot \hat{\mathbf{r}}_{\alpha} - i \varphi_{\mathbf{k}}} \right]. \label{eq-p-3}
\end{equation}
It can be shown numerically (and argued analytically) that, at the lowest energies, $\varepsilon' \rightarrow 0$ and $\varepsilon_{\mathbf{k}} \rightarrow 0$, the leading-order behavior of the two-fermion matrix element in Eq.~(\ref{eq-p-3}) is given by $\langle \phi_{\mathbf{0}}^{\alpha} | \psi_{\varepsilon', -}^{\dag} \, \psi_{\varepsilon_{\mathbf{k}}, +}^{\dag} | \omega \rangle \sim (\varepsilon' + \varepsilon_{\mathbf{k}})^{-1}$. Therefore, the schematic form of $A (\varepsilon_{\mathbf{k}})$ around $\varepsilon_{\mathbf{k}} = 0$, corresponding to the K point of the Brillouin zone, becomes $A (\varepsilon_{\mathbf{k}}) \sim \int_0^{\Lambda} d\varepsilon' g(\varepsilon') [\varepsilon' + \varepsilon_{\mathbf{k}}]^{-1} \sim A_0 + A_1 \varepsilon_{\mathbf{k}} \ln \varepsilon_{\mathbf{k}}$, where $g(\varepsilon) \sim \varepsilon$ is the low-energy matter-fermion density of states. Consequently, while $A (\varepsilon_{\mathbf{k}})$ is finite at $\varepsilon_{\mathbf{k}} = 0$, it is not analytic because its derivative diverges. Moreover, the factor $e^{-i \varphi_{\mathbf{k}}}$ has a nontrivial phase winding around the K point. This nonanalytic behavior of the hybridization function $P_{\mathbf{k},\alpha}$ reflects the gapless Dirac cone of the matter fermions. However, the matter fermions are known to be gapped out by an infinitesimally small magnetic field. Thus, a finite field should remove the nonanalytic behavior by generating an exponential decay for the real-space hybridization function $p_{\mathbf{R},\alpha}$. On a phenomenological level, we can account for this exponential decay by multiplying $p_{\mathbf{R},\alpha}$ with $\exp(-|\mathbf{R}|/\xi)$, which is equivalent to taking a convolution between $P_{\mathbf{k},\alpha}$ and a Lorentzian function of width $\sim 1/\xi$ in momentum space. We emphasize that, while this regularization procedure is important for producing the correct field dependence of the gap opening at the K point, it has negligible effects on all the other results of this work. In practice, we take $\xi=25$ for the hybridization decay length.

To determine the hopping parameter $q$, we match the flux-pair hopping term in Eq.~(3) of the main text with the microscopic model by considering the matrix element of the bare Zeeman term, $\mathcal{H}_h = h \sum_{\alpha} \sum_{\mathbf{r}} \sigma_{\mathbf{r}}^{\alpha}$, where $\sigma_{\mathbf{r}}^{\alpha} = i b_{\mathbf{r}}^{\alpha} c_{\mathbf{r}}^{\phantom{x}} = -\frac{i}{2} \sum_{\beta, \gamma} \epsilon_{\alpha \beta \gamma} b_{\mathbf{r}}^{\beta} b_{\mathbf{r}}^{\gamma}$, between the flux-pair eigenstates $| \tilde{\chi}_{\mathbf{0}}^{x} \rangle = | \phi_{\mathbf{0}}^{x} \rangle \otimes [(\chi_{\mathbf{0}}^{x})^{\dag} | 0 \rangle]$ and $| \tilde{\chi}_{\mathbf{0}}^{y} \rangle = | \phi_{\mathbf{0}}^{y} \rangle \otimes [(\chi_{\mathbf{0}}^{y})^{\dag} | 0 \rangle]$ of the pure Kitaev model. The hopping parameter is then found to be
\begin{equation}
q = \frac{i}{h} \langle \tilde{\chi}_{\mathbf{0}}^{x} | \mathcal{H}_h | \tilde{\chi}_{\mathbf{0}}^{y} \rangle = i \langle \tilde{\chi}_{\mathbf{0}}^{x} | \sigma_{\mathbf{0}}^{z} | \tilde{\chi}_{\mathbf{0}}^{y} \rangle + i \langle \tilde{\chi}_{\mathbf{0}}^{x} | \sigma_{\hat{\mathbf{r}}_{z}}^{z} | \tilde{\chi}_{\mathbf{0}}^{y} \rangle = \langle \phi_{\mathbf{0}}^{x} | \phi_{\mathbf{0}}^{y} \rangle + \langle \phi_{\mathbf{0}}^{x} | i c_{\mathbf{0}}^{\phantom{x}} c_{\hat{\mathbf{r}}_{z}}^{\phantom{x}} | \phi_{\mathbf{0}}^{y} \rangle \equiv q_0 + q^{\prime}. \label{eq-t-1}
\end{equation}
Also, we can straightforwardly determine the signs of $q_0$ and $q^{\prime}$ by considering appropriate products of the corresponding matrix elements $\langle \tilde{\chi}_{\mathbf{0}}^{x} | \sigma_{\mathbf{0}}^{z} | \tilde{\chi}_{\mathbf{0}}^{y} \rangle = -i q_0$ and $\langle \tilde{\chi}_{\mathbf{0}}^{x} | \sigma_{\hat{\mathbf{r}}_{z}}^{z} | \tilde{\chi}_{\mathbf{0}}^{y} \rangle = -i q^{\prime}$ (along with their cyclic permutations in $x,y,z$):
\begin{eqnarray}
P_1 &=& \langle \tilde{\chi}_{\mathbf{0}}^{y} | \sigma_{\mathbf{0}}^{x} | \tilde{\chi}_{\mathbf{0}}^{z} \rangle \langle \tilde{\chi}_{\mathbf{0}}^{z} | \sigma_{\mathbf{0}}^{y} | \tilde{\chi}_{\mathbf{0}}^{x} \rangle \langle \tilde{\chi}_{\mathbf{0}}^{x} | \sigma_{\mathbf{0}}^{z} | \tilde{\chi}_{\mathbf{0}}^{y} \rangle = i q_0^3, \nonumber \\
P_2 &=& \langle \tilde{\chi}_{\mathbf{0}}^{y} | \sigma_{\mathbf{0}}^{z} | \tilde{\chi}_{\mathbf{0}}^{x} \rangle \langle \tilde{\chi}_{\mathbf{0}}^{x} | \sigma_{\hat{\mathbf{r}}_{z}}^{z} | \tilde{\chi}_{\mathbf{0}}^{y} \rangle = q_0 q^{\prime}. \label{eq-t-2}
\end{eqnarray}
Since the individual matrix elements are expected to be ${\cal O}(1)$ due to the absence of an orthogonality catastrophe, it is a reasonable approximation to neglect the projectors to the intermediate states. In this approximation, the products in Eq.~(\ref{eq-t-2}) become
\begin{eqnarray}
P_1 &\approx& \langle \tilde{\chi}_{\mathbf{0}}^{y} | \sigma_{\mathbf{0}}^{x} \sigma_{\mathbf{0}}^{y} \sigma_{\mathbf{0}}^{z} | \tilde{\chi}_{\mathbf{0}}^{y} \rangle = i, \nonumber \\
P_2 &\approx& \langle \tilde{\chi}_{\mathbf{0}}^{y} | \sigma_{\mathbf{0}}^{z} \sigma_{\hat{\mathbf{r}}_{z}}^{z} | \tilde{\chi}_{\mathbf{0}}^{y} \rangle = \langle \sigma_{\mathbf{0}}^{z} \sigma_{\hat{\mathbf{r}}_{z}}^{z} \rangle < 0. \label{eq-t-3}
\end{eqnarray}
From a comparison between Eqs.~(\ref{eq-t-2}) and (\ref{eq-t-3}), we conclude that $q_0$ is positive while $q^{\prime}$ is negative.

\section{Expectation value of the flux operator}

Here we describe how the expectation value of the $\mathbb{Z}_2$ gauge-flux operator $W_p$ in Eq.~(2) of the main text can be computed for the effective Hamiltonian $\tilde{\mathcal{H}}$ in Eq.~(3) in the main text. We first recognize that the dressed bond-fermion operators $\tilde{\chi}_{\mathbf{r}}^{\alpha}$ and the bare bond-fermion operators $\chi_{\mathbf{r}}^{\alpha}$ have exactly the same effect on the flux degrees of freedom as they only differ in an appropriate distortion of the matter-fermion state. Therefore, in terms of the dressed bond variables $\tilde{u}_{jj'}^{\alpha} = i \tilde{b}_{j}^{\alpha} \tilde{b}_{j'}^{\alpha}$ and the ground state $\rvert \tilde{\Omega} \rangle$ of the quadratic fermion Hamiltonian $\tilde{\mathcal{H}}$, the expectation value of the flux operator $W_p$ becomes
\begin{equation}
\langle W_p \rangle = \langle \tilde{\Omega} \rvert \tilde{u}_{12}^{z}\tilde{u}_{32}^{x}\tilde{u}_{34}^{y}\tilde{u}_{54}^{z}\tilde{u}_{56}^{x}\tilde{u}_{16}^{y} \rvert \tilde{\Omega} \rangle = -\langle \tilde{\Omega} \rvert (\tilde{b}_{1}^{z} \tilde{b}_{2}^{z} ) (\tilde{b}_{3}^{x} \tilde{b}_{2}^{x}) (\tilde{b}_{3}^{y} \tilde{b}_{4}^{y}) (\tilde{b}_{5}^{z} \tilde{b}_{4}^{z}) (\tilde{b}_{5}^{x} \tilde{b}_{6}^{x}) (\tilde{b}_{1}^{y} \tilde{b}_{6}^{y}) \rvert \tilde{\Omega} \rangle,
\end{equation}
where the subscript $j = 1,2,...,6$ labels the six sites around the hexagon $p$ (see Fig.~1 of the main text). In turn, this expectation value can be computed by means of Wick's theorem, which reduces the $12$-fermion expectation value to products of two-fermion expectation values, $\langle \tilde{b}_{\bm{r}}^{\alpha} \tilde{b}_{\bm{r}'}^{\alpha'} \rangle = \langle \tilde{\Omega} \rvert \tilde{b}_{\bm{r}}^{\alpha} \tilde{b}_{\bm{r}'}^{\alpha'} \rvert \tilde{\Omega} \rangle$. If we then write the bond-fermion operators $\tilde{b}_{\bm{r}}^{\alpha}$ in terms of the fermion eigenmodes $\gamma_{n, \bm{k}}$ (with $n=1,2,...,8$) of the effective Hamiltonian,
\begin{equation}
\tilde{b}_{\bm{r}\in\nu}^{\alpha} = \frac{1}{\sqrt{N}}\sum_{n,\bm{k}} \left[ (u_{n,-\bm{k}}^{\alpha,\nu})^{*}\gamma_{n,\bm{k}}^{\phantom{\dagger}}+u_{n,\bm{k}}^{\alpha,\nu}\,\gamma_{n,-\bm{k}}^{\dagger} \right] e^{i\bm{k}\cdot\bm{r}}, \quad \,\, \nu=A,B,
\end{equation}
where $u_{n,\bm{k}}^{\alpha,\nu}$ are obtained from a straightforward diagonalization of $\tilde{\mathcal{H}}$, each two-fermion expectation value takes the form
\begin{equation}
\langle \tilde{b}_{\bm{r}\in\nu}^{\alpha} \tilde{b}_{\bm{r}'\in\nu'}^{\alpha'}\rangle = \frac{1}{N} \sum_{n,\bm{k}} (u_{n,\bm{k}}^{\alpha, \nu})^{*} u_{n,\bm{k}}^{\alpha', \nu'} e^{i\bm{k}\cdot(\bm{r}'-\bm{r})}.
\end{equation}
While the ground state $\rvert \tilde{\Omega} \rangle$ contains no bond fermions for $h = 0$, corresponding to $\langle W_p \rangle = 1$, the hybridization between the bond fermions and the matter fermions leads to a finite density of bond fermions for $h > 0$, which corresponds to $\langle W_p \rangle < 1$.

\section{Coefficients of the effective field theory}

Here we provide the coefficients of the effective field theory in Eq.~(11) of the main text. These coefficients can be computed by projecting the effective Hamiltonian in Eq.~(4) of the main text to the two low-energy fermion bands corresponding to Eq.~(9) of the main text. 
For concreteness, the momentum ${\bm k}\equiv (k_x,k_y)$ is described in Cartesian coordinates defined by the unit vectors ${\bf e}_x \parallel \hat{ {\bf r}}_y - \hat{ {\bf r}}_x$ and ${\bf e}_y \parallel \hat{ {\bf r}}_z$ [see Fig.~1 of the main text for definitions of $\hat{ {\bf r}}_{\alpha}$], while the length unit is taken as the lattice constant $a$ (i.e., the distance between two neighboring $A$ sites). 

From a long-wavelength expansion around the $\Gamma$ point, we obtain the following analytical expressions:
\begin{align}
& c_0' = -{\Delta_{\chi}  \over 6+ \Delta_{\chi}} \, {12 \over h_c}, \nonumber \\
& c_z = {\Delta_{\chi} \over 6+\Delta_{\chi}} \left[ 3\left( C \over U \right)^2 {1\over 1-(h_c/h_c')^2} - {3\over 4} - {12V\over U} \right], \nonumber \\
& c_{xx} = -{\Delta_{\chi} \over 2(6+\Delta_{\chi})} \, {C \over U} \, {h_c/h_c' \over 1-(h_c/h_c')^2}, \label{eq-c} \\
& c_{yy} = -{\Delta_{\chi} \over 2(6+\Delta_{\chi})} \, {C \over U} \, {(h_c/h_c')^2 \over 1-(h_c/h_c')^2}, \nonumber \\
& c_{xy} = c_{yx} = 0, \nonumber
\end{align}
where $h_c=\sqrt{6\Delta_{\chi}}/U$ and $h_c'=\Delta_{\chi}/(2\sqrt{3}q)$ are the two critical fields corresponding to the $\Gamma$ point [see the main text], $C=\langle \phi_{\mathbf{0}}^{\alpha} | \omega \rangle$ [see Eq.~\eqref{eq-p-3}], while $U$ and $V$ are defined by $(2/\sqrt{3})\sum_{\alpha}P_{{\bf k},\alpha} = U + V k^2 + {\cal O}(k^3)$. For a finite honeycomb lattice with $N = 121 \times 121$ unit cells, we numerically obtain $C/U \simeq 0.3507$, $h_c/h_c'\simeq 0.3284$, and $V/U \simeq -0.0532$. Therefore, the coefficients in Eq.~\eqref{eq-c} are found to be $c_0' \simeq -1.00$, $c_z \simeq 0.0125$, $c_{xx} \simeq -0.00268$, and $c_{yy} \simeq -0.00088$.

\section{Nonanalytic behavior of the ground-state energy}

Here we analyze the nonanalytic behavior of the ground-state energy $E_G (h)$ at the critical field $h=h_c$. Specifically, we show that the second derivative, $E^{\prime\prime}_G = d^2E_G / d h^{2}$, is discontinuous at $h=h_c$, and provide an expression for its discontinuity, $\Delta E_{G}^{\prime \prime}$, in terms of the effective field theory [see Eqs.~(8) and (11) of the main text]. Given the infrared nature of the singularity, it is useful to write the ground-state energy as a sum of two contributions, $E_G = {\tilde E}_{G} (\Lambda) + \delta E_G (\Lambda)$, which correspond to the long-wavelength modes with momentum $k \equiv \, \rvert {\bf{k}} \rvert < \Lambda$ and the remaining modes with momentum $k > \Lambda$, respectively:
\begin{equation}
{\tilde E}_G(\Lambda)=-\int_{k < \Lambda} 
\frac{d^{2}{\bf{k}}}{\Omega} \,  {\omega_{{\bf{k}}} \over 2}, \quad \delta E_G (\Lambda) = -\int_{\text{BZ}, k > \Lambda} \frac{d^{2}{\bf{k}}}{\Omega} \, {\omega_{{\bf{k}}} \over 2},
\end{equation}
where $\Omega=8\pi^{2}/\sqrt{3}$ is the area of the Brillouin zone (in units of $a^{-2}$), $\Lambda \ll 1$ is an arbitrary cutoff, and $\omega_{{\bf{k}}}$ is the energy of the mode at momentum ${\bf{k}}$ [see Eq.~(12) of the main text]. The second contribution $\delta E_G (\Lambda)$ is analytic at $h=h_c$ because all of its derivatives are well defined. Therefore,
\begin{equation}
\Delta E_{G}^{\prime \prime} = \lim_{h \rightarrow h_c^-}E^{\prime\prime}_G - \lim_{h \rightarrow h_c^+}E^{\prime\prime}_G = \lim_{\Lambda \rightarrow 0} \, \left[ \lim_{h \rightarrow h_c^-} {\tilde E}^{\prime\prime}_G (\Lambda) - \lim_{h \rightarrow h_c^+} {\tilde E}^{\prime\prime}_G (\Lambda) \right].
\end{equation}
In other words, the discontinuity in $E^{\prime\prime}_G$ at the critical field can be completely extracted from the first contribution in the $\Lambda \rightarrow 0$ limit (i.e., the effective field theory) because it does not depend on the cutoff $\Lambda$. 

From Eq.~(12) of the main text, the second derivative of ${\tilde E}_G(\Lambda)$ with respect to the field $h$ is given by 
\begin{equation}
{\tilde E}^{\prime\prime}_G(\Lambda)=-\int_{k<\Lambda} \frac{d^{2}{\bf{k}}}{ \Omega } \left\{ \frac{1}{ 2 \omega_{{\bf{k}}}} \sum_{\mu=x,y,z} \left[ \left(\frac{d\beta_{\bf{k}}^{\mu}}{dh}\right)^{2} +\beta_{{\bf{k}}}^{\mu}\frac{d^{2}\beta_{\bf{k}}^{\mu}}{dh^{2}}\right] - \frac{1}{2\omega_{{\bf{k}}}^3}  \left( \sum_{\mu=x,y,z} \beta_{{\bf{k}}}^{\mu}\frac{d\beta_{\bf{k}}^{\mu}}{dh} \right)^2 \right\}.
\end{equation}
In general, the functions $\beta_{{\bf{k}}}^{\mu}$ depend on the field $h$ via the coefficients $c_0$, $c_z$, and $c_{\eta \nu}$ in Eq.~(11) of the main text. However, in the $\Lambda \rightarrow 0$ limit, the field derivatives of $c_z$ and $c_{\eta \nu}$ are necessarily subdominant with respect to the field derivatives of $c_0$ as they are multiplied by small factors $k^2 < \Lambda^2$ and $k^3 < \Lambda^3$, respectively. Therefore, we focus exclusively on the field dependence of the function $\beta_{{\bf{k}}}^{z}$ via the coefficient $c_0$. If we expand this coefficient around $h=h_c$ as $c_0 (h) = c_0^{\prime} (h-h_c) + \frac{1}{2} c_0^{\prime\prime} (h-h_c)^2 + {\cal O} (h-h_c)^3$, the second derivative of ${\tilde E}_G(\Lambda)$ close to the critical field ($h \simeq h_c$) becomes
\begin{equation}
{\tilde E}^{\prime\prime}_G(\Lambda) = -\int_{k<\Lambda}\frac{d^{2}{\bf{k}}}{ \Omega} \, \frac{\left(\beta_{\bf{k}}^{x}\right)^{2}+\left(\beta_{\bf{k}}^{y}\right)^{2}}{ 2\omega_{{\bf{k}}}^{3}}\left(c_{0}^{\prime}\right)^{2} - \int_{k<\Lambda}\frac{d^{2}{\bf{k}}}{ \Omega } \, \frac{\beta_{{\bf{k}}}^{z}}{ 2 \omega_{\bf{k}}} \, c_{0}^{\prime\prime} \equiv \left[ {\tilde E}^{\prime\prime}_G(\Lambda) \right]^{(1)} + \left[ {\tilde E}^{\prime\prime}_G(\Lambda) \right]^{(2)}. \label{eq-EG}
\end{equation}
We first recognize that the second integral in Eq.~\eqref{eq-EG} vanishes in the $\Lambda \rightarrow 0$ limit for both $h \rightarrow h_c^+$ and $h \rightarrow h_c^-$. Indeed, since $\rvert \beta_{{\bf{k}}}^{z} / \omega_{\bf{k}} \rvert \leq 1$, the magnitude of this integral has an ${\cal O} (\Lambda^2)$ upper bound:
\begin{equation}
\left\rvert \left[ {\tilde E}^{\prime\prime}_G(\Lambda) \right]^{(2)} \right\vert = \left\rvert \int_{k<\Lambda}\frac{d^{2}{\bf{k}}}{ \Omega } \, \frac{\beta_{{\bf{k}}}^{z}}{2\omega_{\bf{k}}} \, c_{0}^{\prime\prime} \right\rvert \leq \int_{k<\Lambda}\frac{d^{2}{\bf{k}}}{ \Omega } \, {\rvert c_{0}^{\prime\prime} \rvert \over 2} = \frac{ \sqrt{3} \Lambda^2 \rvert c_{0}^{\prime\prime} \rvert} { 16 \pi}. \label{eq-result-1}
\end{equation}
In contrast, the first integral in Eq.~\eqref{eq-EG} has completely different behaviors for $h \rightarrow h_c^+$ and $h \rightarrow h_c^-$. To analyze this integral, it is helpful to use polar coordinates, ${\bf{k}}=(k \cos\theta, k \sin\theta)$, write $\beta_{\bf{k}}^{\eta = x,y} = k^{3} \sum_{\nu=x,y} c_{\eta\nu} f_{\nu}(\theta)$ in terms of $f_{x}(\theta)=\cos\theta(3\sin^{2}\theta-\cos^{2}\theta)$ and $f_{y}(\theta)=\sin\theta(3\cos^{2}\theta-\sin^{2}\theta)$, and introduce the positive-definite angular function
\begin{equation}
F(\theta) = k^{-6} \left [\left(\beta_{\bf{k}}^{x}\right)^{2}+\left(\beta_{\bf{k}}^{y}\right)^{2} \right]= \sum_{\eta,\nu=x,y}\left[c_{\eta\nu}f_{\nu}(\theta)\right]^{2}.
\end{equation}
Using $\beta_{\bf{k}}^z \simeq \rvert c_0^{\prime} \rvert (h_c-h) + c_z k^2$ (where $c_0^{\prime} < 0$ and $c_z > 0$), the first integral in Eq.~\eqref{eq-EG} can then be written as
\begin{equation}
\left[ {\tilde E}^{\prime\prime}_G(\Lambda) \right]^{(1)} = -{\sqrt{3} \over 4}\left(\frac{c_{0}^{\prime}} {2\pi} \right)^2 
\int_{0}^{2\pi} d\theta \int_{0}^{\Lambda} dk \, \frac{k^7 F(\theta)} {\left\{ k^6 F(\theta) + \left[ \rvert c_0^{\prime} \rvert (h_c-h) + c_z k^2 \right]^2 \right\}^{3/2}}. \label{eq-int}
\end{equation}
Below the critical field, $h < h_c$, the denominator of the integrand is bounded from below by $c_z^3 k^6$. Thus, the integral vanishes in the $\Lambda \rightarrow 0$ limit for $h \rightarrow h_c^-$ because its magnitude has an ${\cal O} (\Lambda^2)$ upper bound:
\begin{equation}
\left\rvert \lim_{h \rightarrow h_c^-} \left[ {\tilde E}^{\prime\prime}_G(\Lambda) \right]^{(1)} \right\rvert \leq  {\sqrt{3} \over 4}  \left( \frac{c_{0}^{\prime}} {2\pi} \right)^2 \int_{0}^{2\pi} d\theta \int_{0}^{\Lambda} dk \, \frac{k F_{\max}} {c_z^3} = \frac{ \sqrt{3} \Lambda^2 \left( c_{0}^{\prime} \right)^2 F_{\max}} {16 \pi c_z^3}, \label{eq-result-2}
\end{equation}
where $F_{\max} = \max_{\theta} F(\theta)$. In contrast, above the critical field, $h > h_c$, the denominator of the integrand in Eq.~\eqref{eq-int} is very small along the low-energy ring of radius $K = \sqrt{\rvert c_0^{\prime} \rvert (h-h_c) / c_z}$. Expanding the integrand around this radius $K$ by introducing a new radial variable, $x = (k - K) / K^2$, the integral in Eq.~\eqref{eq-int} then becomes
\begin{equation}
\left[ {\tilde E}^{\prime\prime}_G(\Lambda) \right]^{(1)} = - {\sqrt{3} \over 4} \left( \frac{c_{0}^{\prime}} {2\pi} \right)^2 \int_{0}^{2\pi} d\theta \int_{-1/K}^{(\Lambda-K)/K^2} dx \, \frac{K^9 (1 + Kx)^7 F(\theta)} {\left\{ K^6 (1 + Kx)^6 F(\theta) + \left[ -c_z K^2 + c_z K^2 (1 + Kx)^2 \right]^2 \right\}^{3/2}}.
\end{equation}
Next, in the limit of $h \rightarrow h_c^+$, corresponding to $K \rightarrow 0$, the integral reduces to
\begin{equation}
\lim_{h \rightarrow h_c^+} \left[ {\tilde E}^{\prime\prime}_G(\Lambda) \right]^{(1)} = - {\sqrt{3} \over 4}
\left( \frac{c_{0}^{\prime}} {2\pi} \right)^2 \int_{0}^{2\pi} d\theta \int_{-\infty}^{+\infty} dx \, \frac{F(\theta)} {\left[ F(\theta) + 4 c_z^2 x^2 \right]^{3/2}} = -\frac{ \sqrt{3} \left( c_{0}^{\prime} \right)^2} {8\pi c_z}. \label{eq-result-3}
\end{equation}
This $K$-independent result has a simple interpretation: the first integral in Eq.~\eqref{eq-EG} is dominated by an annulus of radius $K$ and width ${\cal O} (K^2)$ in which the integrand is ${\cal O} (K^{-3})$. Finally, by collecting the results from Eqs.~\eqref{eq-result-1}, \eqref{eq-result-2}, and \eqref{eq-result-3}, we conclude that the discontinuity in the second derivative of the ground-state energy at the critical field is given by
\begin{equation}
\Delta E_{G}^{\prime \prime} = \lim_{h \rightarrow h_c^-}E^{\prime\prime}_G - \lim_{h \rightarrow h_c^+}E^{\prime\prime}_G = \lim_{\Lambda \rightarrow 0} \, \left[ \lim_{h \rightarrow h_c^-} {\tilde E}^{\prime\prime}_G (\Lambda) - \lim_{h \rightarrow h_c^+} {\tilde E}^{\prime\prime}_G (\Lambda) \right] = \frac{\sqrt{3}\left( c_{0}^{\prime} \right)^2} {8\pi c_z}.
\end{equation}
Remarkably, the discontinuity only depends on two parameters, $c_0^{\prime}$ and $c_z$, of the effective field theory.

\end{document}